# Viterbi Extraction tutorial with Hidden Markov Toolkit

Zulkarnaen Hatala [1], Victor Puturuhu [2]
Electrical Engineering Department
*Politeknik Negeri Ambon*
Ambon, Indonesia
[1]dzulqarnaenhatala@gmail.com
[2]victorputuruhu@gmail.com

***Abstract*—An algorithm used to extract HMM parameters is revisited. Most parts of the extraction process are taken from implemented Hidden Markov Toolkit (HTK) program under name HInit. The algorithm itself shows a few variations compared to another domain of implementations. The HMM model is introduced briefly based on the theory of Discrete Time Markov Chain. We schematically outline the Viterbi method implemented in HTK. Iterative definition of the method which is ready to be implemented in computer programs is reviewed. We also illustrate the method calculation precisely using manual calculation and extensive graphical illustration. The distribution of observation probability used is simply independent Gaussians r.v.s. The purpose of the content is not to justify the performance or accuracy of the method applied in a specific area. This writing merely to describe how the algorithm is performed. The whole content should enlighten the audience the insight of the Viterbi Extraction method used by HTK.**

## I. Introduction

Hidden Markov Toolkit (HTK) [1] is state of the art in speech recognition fields. HTK is released with its source code open that provide advantage to all researchers specially beginners and young scientists who want to study the implementation of ASR. Even its source code is available but HTK is considerable a big project with thousands of C lines. An effort must be done in order to use the libraries of HTK and to extend its functionalities. Thus, by studying directly into its C code intensively, researcher can effectively apply HTK to a new language. Or even correcting errors encounter while applying. Looking deep inside into existing C code of HTK is mandatory to master and understand its data structures and code sequences. HTK is an integrated suite of software tools for building and manipulating continuous density Hidden Markov Models (HMMs). It consists of a set of library modules and a set of tools (executable programs). HTK is written in ANSI C and runs any modern OS. It is currently used in over many speech laboratories around the world and it is also used for teaching in number of Universities.

## II. Definitions and Terms

### A. Hidden Markov Model

HMM[2] is a discrete time Markov chain (DTMC) [3] that emit an observation when sitting in one of its states. An HMM is completely defined by the initial and transition probability of Markov chain (MC) and state's observation condition probabilities. The distribution of observation only depends on the current state of Markov chain. In some application not all states of Markov chain emit observations. Set of states that emit an observation is called emitting states, while non emitting states could be an entry or start state, absorption or exiting state and intermediate or delaying state. The entry state is the Markov chain state with initial probability 1.

One application of Hidden Markov Model is in speech recognition. The observations of HMM are used to model sequence of speech feature vectors like Mel Frequency Cepstral Coefficients (MFCC) [4]. MFCC is assumed to be generated by emitting hidden states of the underlying DTMC. According to [2] an HMM is completely determined by parameter set:

$$\boldsymbol{\lambda} = (\mathbf{A}, \mathbf{B}, \boldsymbol{\pi}) \tag{1}$$

Where **A** and $\boldsymbol{\pi}$ is the transition probability matrix and initial probability vector of DTMC and **B** is the observation probability.

### B. Viterbi algorithm

Baum Viterbi algorithm is used to estimate parameters of HMM in (1). This method is known in literature as Viterbi Training, Viterbi segmentation, Baum Viterbi, segmental K-means [5], classification EM, hard EM, MAP path estimator [6], etc. The convergence properties of this algorithm is also elaborated in [6].

As taken from [2], the iterative algorithm for finding the best state sequence is:

*1) Initialization,* set scores from initial state into states that generate first observation:

$$\delta_1(j) = \pi_j b_j(O_1) \tag{2}$$

*2) Recursion, propagating next observation and finding maximum score for each destination state j:*

$$\delta_t(j) = \left[\max_i \delta_{t-1}(i) a_{ij}\right] b_j(O_t), \tag{3}$$

And save state entries in backtrack matrix:

$$\psi_t(j) = \arg\max\left[\max_i \delta_{t-1}(i) a_{ij}\right] \tag{4}$$

$$2 \le t \le T-1, \;\; 1 \le j \le S$$

*3) Termination, finding best score and best path end state:*

$$P^* = \max_i \left[ \delta_T(i) \right] \tag{5}$$

$$q_T^* = \arg\max_i \left[ \delta_T(i) \right] \tag{6}$$

*4) Backtraking, recursively align path sequence from best path end state:*

$$q_t^* = \psi_{t+1}\left(q_{t+1}^*\right),\ t = T-1, T-2, \ldots, 1 \tag{7}$$

The time complexity of the algorithm for a fixed number of states and *T* number of observations are *O(T)*. The memory requirement for this algorithm has size of *T×S*.

## *C. Hidden Markov Toolkit*

The components of HTK are depicted in figure 1. In the outer block outlined HTK libraries which were written in ANSI C. In the central block, HTK Tools are ready to use programs such as HInit, HCopy, HRest, HVite and so on.

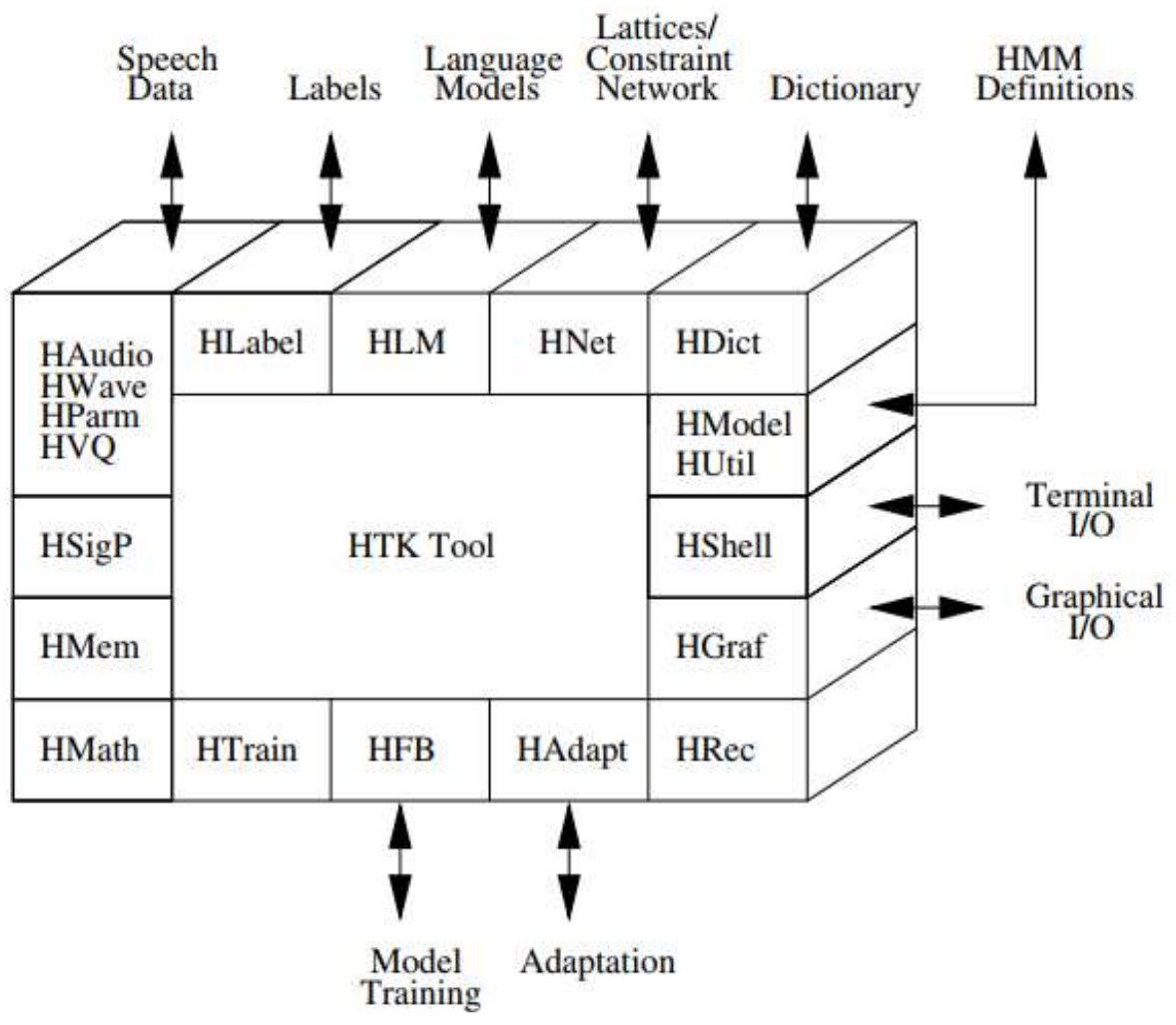

Fig. 1. Components of HTK

HTK source code which is provided freely open contains libraries that consist of predefined C language data structures, functions definitions and implementations. The source code also contains ready to use executable programs that can be run in Operating System.

# III. Illustration

To illustrate Viterbi method, let's consider a model of HMM that can be used in HTK. The underlying DTMC is depicted in figure-2. This is a four states HMM with a start state 1, two emitting states 2, 3 and an absorption state 4. This HMM is used to model a sequence of MFCC associated with single subword or phonem in speech recognition. For example the Indonesian digit *satu* (english: *one*) can be broken into its subwords as *s-ah-t-uh*. In this case subword *s* has its own HMM definition, so are *ah, t* and *uh*. This HMM has property that the initial probability of state 1 is 1, and the only transition from state S1 is limited only into state 2.

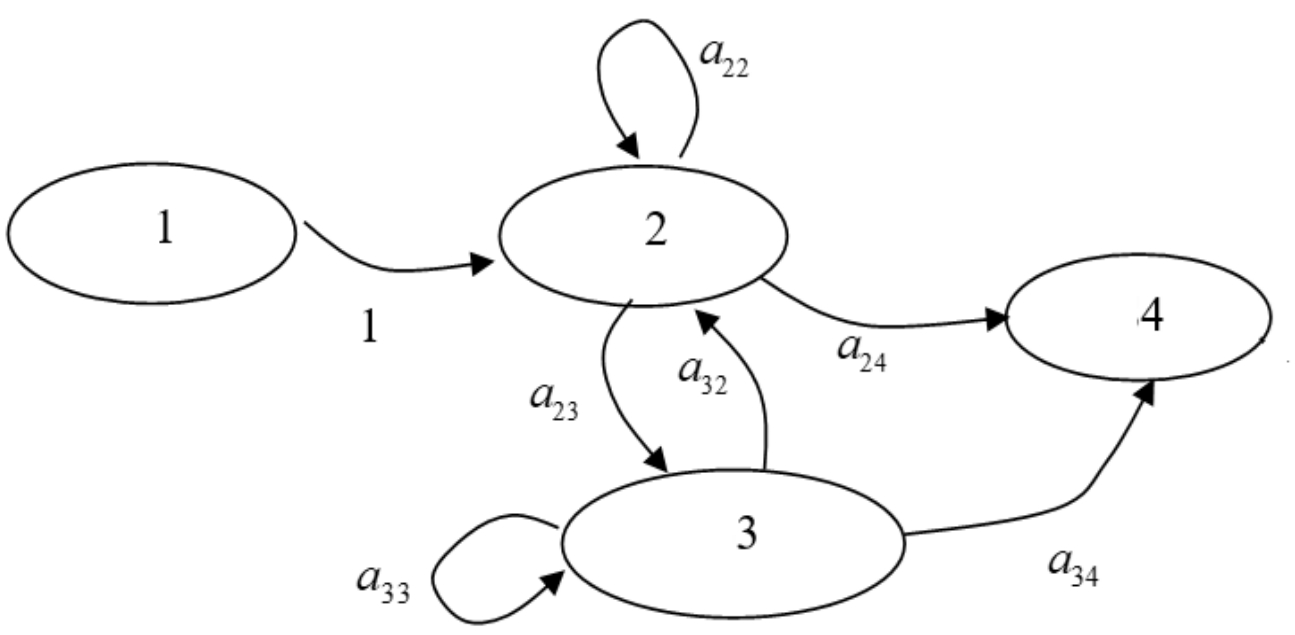

Fig. 2. HTK's HMM model

## *A. Observation data*

Consider observations data in TABLE I. At each time *t=1,2,...,12,* a vector:

$$\mathbf{x} = \left[ x_{t1} \quad x_{t2} \quad x_{t3} \right],\ t = 1, \ldots, 12 \tag{8}$$

is assumed to be generated, either by state 2 or state 3 of DTMC in figure 2.

TABLE I. Observation Data

| *t* | $x_{t1}$ | $x_{t2}$ | $x_{t3}$ |
|---|---|---|---|
| 1 | -1.115696192 | -1.014122963 | -0.244220227 |
| 2 | -0.971390247 | -0.823073566 | -0.661046565 |
| 3 | -0.399597883 | -0.510152936 | -0.782005250 |
| 4 | 0.652983546 | 0.032239955 | -0.676724792 |
| 5 | 1.174029231 | 0.492249459 | -0.531797767 |
| 6 | 1.049691796 | 0.873368561 | -0.658340454 |
| 7 | 0.582641065 | 1.254104257 | -0.705632925 |
| 8 | -0.179997236 | 1.284092903 | -0.751379013 |
| 9 | -0.513338625 | 0.562875986 | -0.750906527 |
| 10 | -0.466798663 | -0.551046491 | -0.449841380 |
| 11 | -0.309872597 | -1.142085671 | -0.048693269 |
| 12 | -0.206728458 | -1.149199367 | 0.414609402 |

Actually these data are taken from 3 last MFCC's coefficients generated by HTK program HCopy operated on a recorded sound file.

## *B. Observation distribution*

Each item in filtered MFCC vector is assumed to be mixture of Gaussian. When there is only single component of Gaussian per items, then the observation probability for MFCC vector is simply the multiplication of the item. In this case the observation parameter in equation (1) which can be written as:

$$\mathbf{B} = \left[ b_2(t) \quad b_3(t) \right] \tag{9}$$

Its logarithm is simply calculated as in (10):

$$\log\left(b_j(t)\right) = -\frac{1}{2}\left( G_j + \sum_i \frac{\left(x_{ti} - \mu_{ji}\right)^2}{\sigma_{ji}^2} \right) \tag{10}$$

$$G_j = 3\log(2\pi) + \sum_i \log \sigma_{ji}^2 , \quad (11)$$

$$j = 2,3, \quad i = 1,2,3, \quad t = 1,2,...,12$$

Equation (10) can also be stated using Matlab notation:

$$\log_bj_t = sum(\log(normpdf(\mathbf{x_t}; \mathbf{mu_j}; \mathbf{var_j}))) \quad (12)$$

*C. HTK program HInit*

In this section we're going to extract parameter set for HMM with DTMC in figure 2, based on observation in TABLE-I. In HTK, Viterbi algorithm is implemented in program called HInit. The operation of HInit is summarized in figure 3.

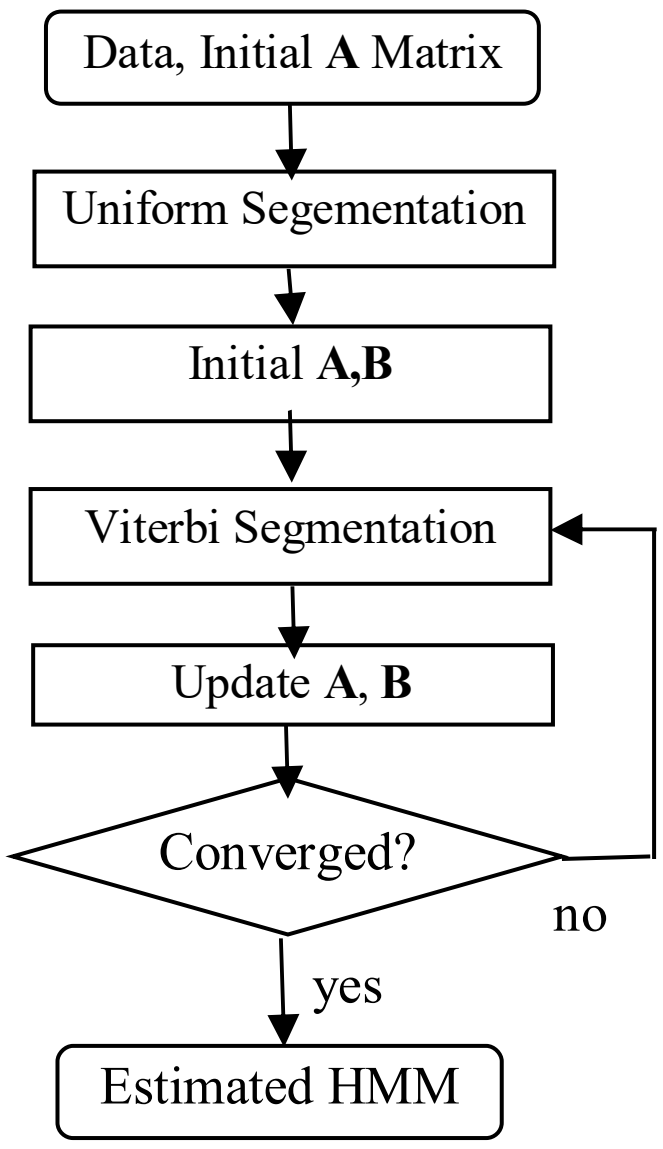

Fig. 3. Viterbi algorithm on HInit

*D. Uniform segmentation and parameter initialization*

The initial probability $\boldsymbol{\pi}$ in (1) is not estimated since at initial time $t$=0 the DTMC on figure 2 always start at state 1 with probability 1. Moving to figure 3, initial transition probability matrix **A** in (1) is needed. Let fix this matrix as in (13).

$$\mathbf{A} = [a_{ij}] = \begin{bmatrix} 0.0 & 1.0 & 0.0 & 0.0 \\ 0.0 & 0.1 & 0.4 & 0.5 \\ 0.0 & 0.8 & 0.1 & 0.1 \\ 0.0 & 0.0 & 0.0 & 0.0 \end{bmatrix} \quad (13)$$

All entries of the 4th row in (13) is zero because state 4 is the absorption state. By uniform segmentation it means that number of observations are assumed to be generated equally between states. Since we have only 2 emitting states for 12 observations, then the first 6 observations are segmented into state 2 and the rest is assumed to be generated by state 3. By this segmentation, observation distribution in (3), means and variances of Gaussians are initialized as in TABLE II:

TABLE II. UNIFORM SEGMENTATION PARAMETERS

| State $j$ | Mean $\hat{\boldsymbol{\mu}}_j$ | Variances $\hat{\boldsymbol{\sigma}}_j^2$ |
|---|---|---|
| 2 | 0.0650 -0.1583 -0.5923 | 0.8717 0.4701 0.0295 |
| 3 | -0.1823 0.0432 -0.3820 | 0.1322 1.0758 0.1880 |

Parameter initialization step is complete since all parameters in (1) is identified. Now the Viterbi method is about to begin.

*E. Viterbi Extraction*

An illustration for the estimation process is presented in figure 4 for time $t=1$ to $t$=7 and in figure 5 for $t$=8 *to* $t$=12. Values inside circles representing delta in (3). Green circles are for state 2 and red circles are for state 3. Black circles are starting and absorption states.

*1) Initialization, Viterbi cost for the first propagation is only for observation generated by state 2.*

$$\delta_1(2) = \log(P(\mathbf{x}_1 \mid q_1 = 2))$$
$$= -4.1817$$

The only backtrack entry exists, is for state 2:

$$\psi_1(2) = 1$$

*2) Recursion, for all next observations from time t=2 until t=12:*

$t$=2: At this time, observation can be generated by staying in state 2 or moving to state 3 with total cost each:

$$\delta_2(2) = -7.1013, \ \delta_2(3) = -8.3930$$

The entries in backtrack matrix each:

$$\psi_2(2) = 2, \ \psi_2(3) = 2$$

t=3: at these times propagating into state 2 or 3 can be reach either from state 2 or 3 also. In these cases, we choose source with the best probability or minimal cost to calculate (3). We remove unsurvival path components $q_2 = 2, q_3 = 2$ and $q_2 = 3, q_3 = 3$. Eliminated path components is shown in figure 4 as a black arrow crossed by a red line. Now we have values for (3) as:

$$\delta_3(2) = -10.59, \ \delta_3(3) = -10.81$$

And values for (4) as:

$$\psi_3(2) = 3, \ \psi_3(3) = 2$$

By iteratively continuing in this fashion we have all values for (3) and (4) for every time index from $t$=3 until $t$=12. Again, for an illustration, the values are drawing inside circles at figure 4 and figure 5.

*3) Termination:* maximum probability or minimum cost when generating last observation at time $t$=12 and also entering exit state 4 is achieved via state 3.

$$P_1^* = \delta_{12}(3) + \log(a_{34})$$
$$= -44.0826$$
$$q_T^* = 3 \quad (14)$$

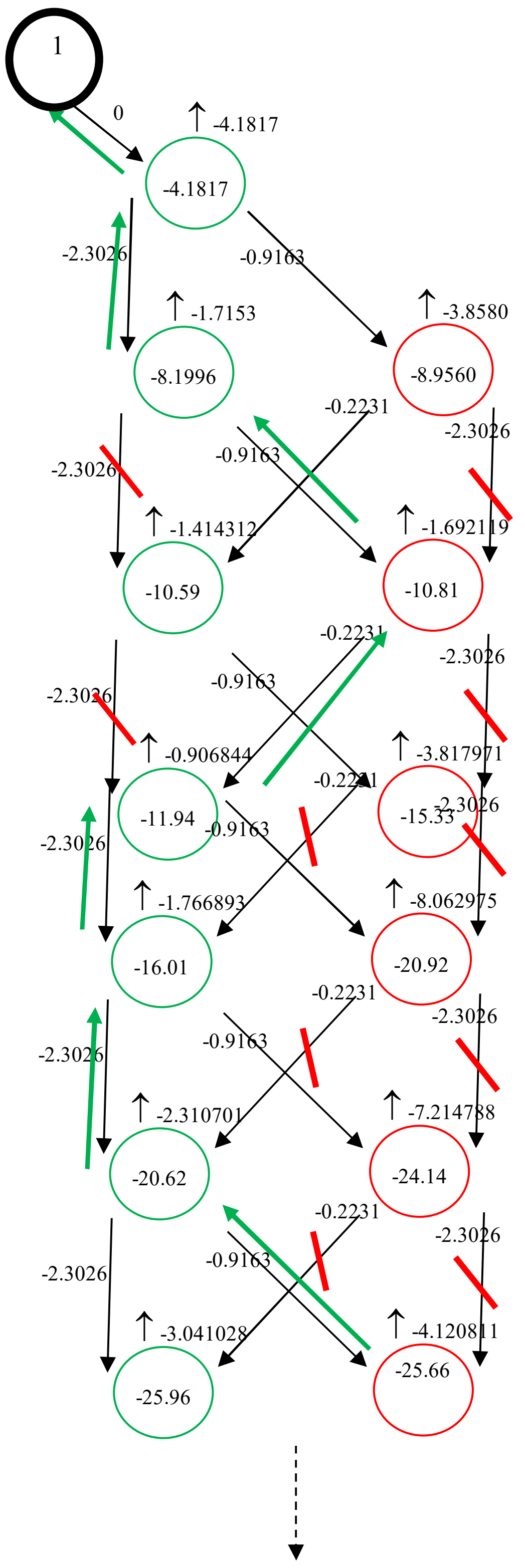

Fig. 4. Viterbi iteration for time $t$=1 to $t$=7

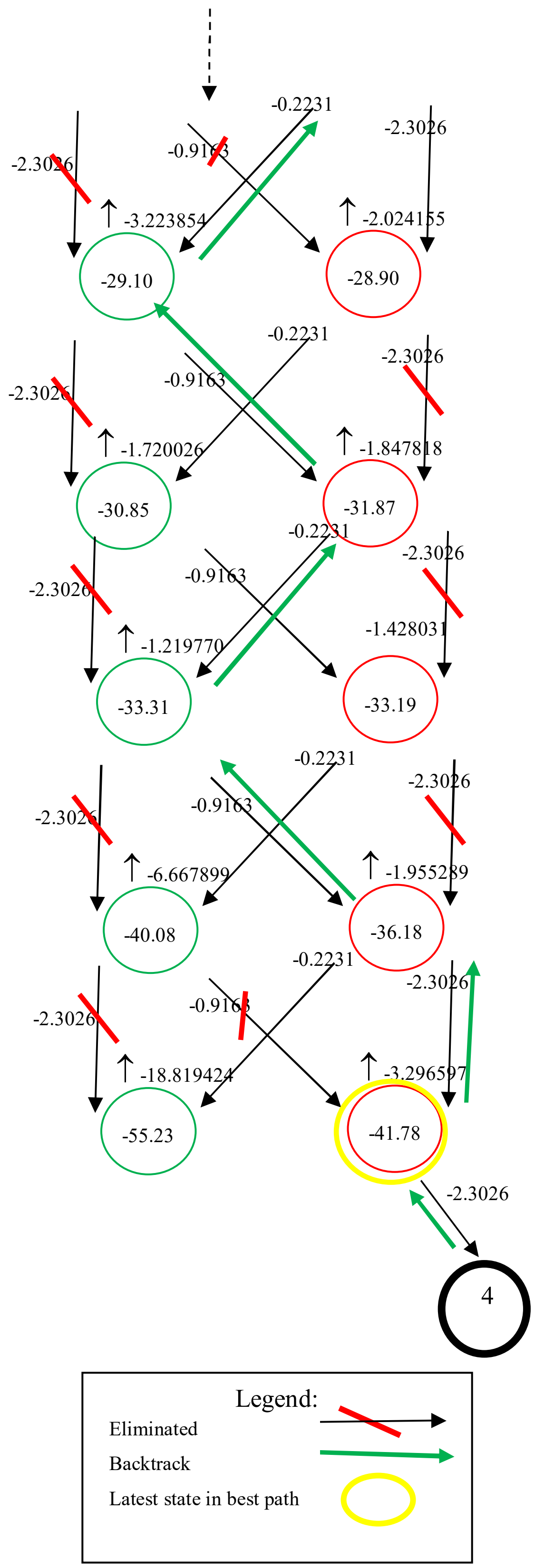

Fig. 5. Viterbi iteration for time $t$=8 to $t$=12

Backtracking: best path sequence can be traced back using (14) following all values inside backtrack matrix generated in step 3). *Backtrack matrix is shown in TABLE III.*

TABLE III. TRACEBACK MATRIX

| t / state | 1 | 2 | 3 | 4 | 5 | 6 | 7 | 8 | 9 | 10 | 11 | 12 | 13 |
|---|---|---|---|---|---|---|---|---|---|---|---|---|---|
| **2** | 1 | **2** | 3 | **3** | **2** | **2** | 2 | **3** | 3 | **3** | 3 | 3 | - |
| **3** | - | 2 | **2** | 2 | 2 | 2 | **2** | 2 | **2** | 2 | **2** | **3** | - |
| **4** | - | - | - | - | - | - | - | - | - | - | - | - | **3** |

Backtracking is illustrated as following green arrows in figure 4 and figure 5. The same goal is achieved by following grey shaded cell in TABLE III. Using these ways, at this iteration we have for (7) our best path sequence of states in table IV.

TABLE IV. STATES SEQUENCE EXTRACTED

| T | 0 | 1 | 2 | 3 | 4 | 5 | 6 | 7 | 8 | 9 | 10 | 11 | 12 | 13 |
|---|---|---|---|---|---|---|---|---|---|---|---|---|---|---|
| STATE | 1 | 2 | 2 | 3 | 2 | 2 | 2 | 3 | 2 | 3 | 2 | 3 | 3 | 4 |

## F. *HMM Parameter Updating*

Using state segmentation results in TABLE IV and observation data in TABLE I, HMM parameters now can be updated. From table IV the transition between state can be counted and be normalized into a new transition probability which is shown in (15)

$$\mathbf{A}_{iter1} = \begin{bmatrix} 0 & 1 & 0 & 0 \\ 0 & \frac{3}{7} & \frac{4}{7} & 0 \\ 0 & \frac{3}{5} & \frac{1}{5} & \frac{1}{5} \\ 0 & 0 & 0 & 0 \end{bmatrix} \qquad (15)$$

Gaussian parameters can be determined the same way TABLE II is calculated. But instead of using uniform segmentation, now segmentation in TABLE IV is used. Updated parameters are shown in TABLE V.

TABLE V. ITERATION-1: GAUSSIAN PARAMETERS

| State $j$ | Mean $\hat{\boldsymbol{\mu}}_j$ | Variances $\hat{\boldsymbol{\sigma}}_j^2$ |
|---|---|---|
| 2 | 0.0204 0.0419 -0.5676 | 0.7633 0.6644 0.0260 |
| 3 | -0.1694 -0.1969 -0.3745 | -0.1516 0.9162 0.2292 |

Since this is the first Viterbi iteration, based on figure 3 we have to perform an iteration at least once again. After the second iteration then we can decide if the whole process is convergent based on (15):

$$\left|P_{k+1}^* - P_k^*\right| \le \varepsilon \qquad (15)$$

In (15), $P_k^*$ is the value of (5) at $k$th iteration. Viterbi process is terminated whenever the probability difference of the two consecutive iterations is less then $\varepsilon$ . The next iterations of Viterbi segmentation is performed the same way which is illustrated in figure 4 and figure 5. That is the parameters estimated from previous iteration are used to segment the states of the next iteration. In our example of HMM with underlying DTMC on figure 2, we will stop at iteration 5-th if:

$$\varepsilon = 0.0001$$

Each iteration values of (15) for our example is presented in TABLE VI.

TABLE VI. ITERATIONS CONVERGENCE VALUES

| Iteration-*k*th. | $P_k^*$ | $P_k^* - P_{k-1}^*$ |
|---|---|---|
| Iteration-1st | -44.08260 | - |
| Iteration-2nd | -35.15588 | 8.9267 |
| Iteration-3th | -29.51328 | 5.6426 |
| Iteration-4th | -18.87455 | 10.6387 |
| Iteration-5th | -18.87455 | Less than 0.0001 |

## IV. DISCUSSIONS

In this tutorial the detail of the Viterbi Algorithm to estimate HMM parameters are presented. The method adopted the implementation used by HTK. The algorithm by HTK may show some variations against another domain of where Viterbi is implemented. For further explanation of this algorithm against its implementation in HTK, the observation data should be more massive and varying. Nonetheless, the main concept is already described and illustrated.

.

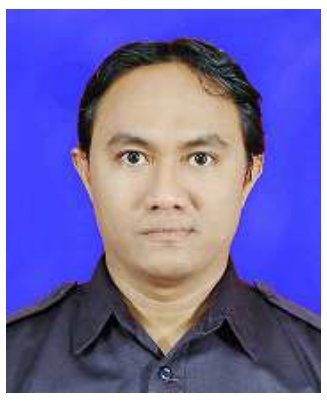

Zulkarnaen Hatala was born in Ambon, on 19 Agustus 1977. He received Sarjana Teknik on Informatics at 2002 and Master Teknik on Telecommunications at 2005. Both degress are from Telkom University, Bandung Indonesia. From 2005 to 2009 he thought computer science and informatics in private universities. Since 2009 he works at Politeknik Negeri Ambon, Indonesia as a lecturer on Electrical Engineering Department.